\documentclass[twocolumn,showpacs,amsmath,amssymb, superscriptaddress]{revtex4}

\usepackage{amsmath}
\usepackage{amssymb}
\usepackage{amsfonts}
\usepackage{graphicx}
\usepackage{float}
\usepackage{color}
\usepackage{amsthm}

\DeclareMathOperator{\Tr}{Tr}

\newcommand{\angl}[1]{{\left\langle #1 \right\rangle}}

\makeatletter
\def\loweq@align#1#2{\lower.6ex\vbox{\baselineskip\z@skip\lineskip\z@
    \ialign{$\m@th#1\hfil##\hfil$\crcr#2\crcr=\crcr}}}
\def\lowsim@align#1#2{\lower.6ex\vbox{\baselineskip\z@skip\lineskip\z@
    \ialign{$\m@th#1\hfil##\hfil$\crcr#2\crcr\sim\crcr}}}
\def\geqq{\mathrel{\mathpalette\loweq@align >}}
\def\leqq{\mathrel{\mathpalette\loweq@align <}}
\def\grsim{\mathrel{\mathpalette\lowsim@align >}}
\def\lesssim{\mathrel{\mathpalette\lowsim@align <}}
\def\gsim{\mathrel{\mathpalette\lowsim@align >}}
\def\lsim{\mathrel{\mathpalette\lowsim@align <}}
\makeatother
\newcommand{\grless} 
{ {\, \raise-.24em\hbox{$<$} \hspace{-0.8em} \raise.31em\hbox{$>$}\, } }
\newcommand{\lessgr} 
{ {\, \raise-.24em\hbox{$>$} \hspace{-0.8em} \raise.31em\hbox{$<$}\, } }
\newfont{\bg}{cmr10 scaled\magstep4}                    
\newcommand{\bigzerou}{\smash{\lower1.7ex\hbox{\bg 0}}}

\newcommand{\crl}[1]{[-\infty,\infty]}

\newcommand{\ket}[1]{|{#1}\rangle}

\newcommand{\Ref}[1]{(\ref{#1})}

\newcommand{\da}[1]{#1^\dag}

\newcommand{\alg}[1]{{\mathfrak #1}}

\newcommand{\av}[1]{\langle#1\rangle}

\begin{document}

\title{Time-optimal Unitary Operations in Ising Chains II: \\ Unequal Couplings and Fixed Fidelity}
  \author{Alberto Carlini}
 \email{acarlini@mfn.unipmn.it}
 \affiliation{Dipartimento di Scienze ed Innovazione Tecnologica, Universita' del Piemonte Orientale, Alessandria, Italy}
 \affiliation{Istituto Nazionale di Fisica Nucleare, Sezione di Torino, Gruppo Collegato di Alessandria, Italy}

\author{Tatsuhiko Koike}
 \email{koike@phys.keio.ac.jp}
 \affiliation{Department of Physics, Keio University, Yokohama, Japan}

\begin{abstract}
We analytically determine the minimal time and the optimal control laws required for the realization, 
up to an assigned fidelity and with a fixed energy available, of entangling quantum gates ($\mathrm{CNOT}$) 
between indirectly coupled qubits of a trilinear Ising chain.
The control is coherent and open loop, and it is represented by a local and continuous magnetic field acting on the intermediate qubit. The time cost of this local quantum operation is not restricted to be zero.
When the matching with the target gate is perfect (fidelity equal to one) we provide exact solutions for the case of equal Ising coupling. For the more general case when some error is tolerated (fidelity smaller than one) we give perturbative solutions for unequal couplings.
Comparison with previous numerical solutions for the minimal time to generate the same gates with the same Ising Hamiltonian but with instantaneous local controls shows that the latter are not time-optimal. 
 
\end{abstract}  


\maketitle

\section{Introduction}

Time-optimal control is an active branch of the wider arena of quantum control.
Quantum control has been applied successfully to a series of physical problems, e.g., molecular dynamics \cite{shapiro},
quantum information \cite{chuangnielsen}-\cite{brif}, multidimensional advanced spectroscopy techniques in 
nuclear magnetic resonance (NMR) experiments \cite{cavanagh}.
Time-optimal quantum computing aims at minimizing the time cost to achieve a certain target.
It may be of relevance both from the theoretical point of view, e.g., in order to determine the fundamental maximal
speed at which quantum information may be transported \cite{margolus}-\cite{murphy}, to find out faster quantum
algorithms and solve complex problems \cite{nielsen}, or it may serve more practical but essential purposes such 
as determining the fastest way to realize a certain unitary gate before the ubiquitous decoherence effects disrupt the
computation, and in general to define a more physical meaning to the complexity of quantum algorithms \cite{schulte}.
The results published in time-optimal quantum control are numerous and it is certainly not the scope of the present paper
to give an accurate or comprehensive review of this fascinating subject. Some literature may be found, e.g., in \cite{ising}.

In a series of papers \cite{pure}-\cite{complexity} we introduced a theoretical framework for time-optimal quantum computing based on the action principle where  the Hamiltonian is subject to a set of constraints (e.g., a finite energy,
certain qubit interactions are forbidden), and we called it the Quantum Brachistochrone (QB, \footnote{From the 
Greek  ``$\beta\rho\alpha\chi\iota\sigma\tau o\zeta$", i.e., fast, and ``$\chi\rho o\nu o\zeta$", i.e time.}).
The QB was applied to quantum state evolution of pure \cite{pure} and mixed states
\cite{mixed}, and to the optimal realization of unitary gates \cite{unitary}.
The predictions of the QB on the time complexity of generating unitaries with Hamiltonians which contains only 
one and two qubit interaction terms can be found in \cite{complexity}. 
More recently, we discussed the problem of optimally generating entangling unitaries in a trilinear Ising chain subject to
a coherent open loop control \cite{ising}. Assuming a fixed energy available and
that local operations on qubits are time-consuming, we found the laws of optimal control and the optimal minimal time 
for realizing a $\mathrm{CNOT}$ gate between indirectly coupled qubits. This time is shorter than previous results
found in the literature so far \cite{khaneja2}.
The importance of the study of quantum gates in physical Hamiltonians with indirect couplings is that non optimized
procedures are usually time costly and easily prone to decoherence effects and a degradation of the gate fidelity,  
both of which are critical for practical quantum computing.
Interactions between indirectly coupled qubits via an intermediate qubit are also a typical scenario in a wide array of promising (scalable) experimental realizations of quantum information processing, e.g., in solid state architectures 
\cite{kane}, crystal lattice architectures \cite{yamaguchi}, superconducting \cite{clarke} and 
NMR technologies \cite{ernst}. 

Here would like to elaborate upon \cite{ising} on the problem of time-optimal generating unitary gates in linear spin chains.
The motivation is essentially twofold.
On the one hand, we would like to introduce a general theoretical framework for finding the time-optimal way of
generating a unitary gate in the more realistic situation where the target can be reached within a finite, tolerable error,
in other words with a fidelity close to but smaller than one.
On the other hand, we would like to extend our previous analysis of Ising models to the case when the interaction couplings
between spins are non equal. 
Work related to our research can be found in \cite{yuan}-\cite{nimbalkar}, where analytical geometrical methods 
\cite{khaneja} and numerical algorithms \cite{grape} are used.
We obtain new and interesting results for the minimal time duration of certain entangling
gates and propose some experiments to test our theoretical formulas.

The paper is organized as follows.
In Section II we briefly review the main features of the QBT formalism for the time-optimal synthesis 
of unitary quantum gates with a given final fidelity.
In Section III we apply the formalism of the QBT and we introduce the main formulas for 
the problem of the efficient generation of the gate $U^s_{13}$ between the indirectly coupled boundary qubits 1 and 3
of a three-linear Ising chain subject to a local, time consuming control on the intermediate qubit, when a finite energy is available.
Section IV is devoted to the discussion of the case of a final perfect matching between the time-optimal unitary evolution and the target, while Section V analyzes the more general case when some errors are tolerated and the final fidelity is
less than one. We show that in general the minimal time required to reach the target if smaller than previous estimates 
given in the literature. We also propose a series of possible experimental situations in which the results of the QBT formalism can be verified.
Section VI is used to discuss the time-optimal generation with fixed fidelity of the gate $\mathrm{CNOT}(1, 3)$,
which can be achieved using a slightly modified interaction Hamiltonian plus the same local operation on qubit 2 and the same energy available.
Finally, Section VII is devoted to the summary and discussion of our results. 
The derivation of the main formulas exposed in the main Sections is outlined in Appendix A-D.

\section{The Quantum Brachistochrone with fixed fidelity}

The goal is to minimize the time $T$ it takes to perform a unitary operation under the condition that the fidelity $f$ 
between the time optimal unitary at time $T$ and a certain target is fixed.
In other words, we want to realize a certain target unitary $U_f$, but we tolerate that the target is reached only approximately, i.e. with a fidelity $f < 1$.
We call this problem the Quantum Brachistochrone with Fidelity (QBF).
We also assume that the Hamiltonian $H$ is controllable and should obey some constraints, dictated either by theoretical conditions (e.g., only certain interactions among qubits are allowed) or by experimental requirements (e.g., a finite energy),
The unitary operator $U(t)$ should then obey the Schr\"odinger equation. 
The QBF problem can be formulated  in terms of the minimization of the following action:
\begin{align}
  \label{eq-action}
  S(U,H, U(T); \alpha, \Lambda,\lambda, \lambda_j)&:=\lambda [|\av{U(T),U_f }|^2-(Nf)^2]
  \nonumber 
  \\ 
  &+\int_0^1 ds \left[ \alpha N +L_S+L_C\right],
    \\
  \label{eq-LS}
  L_S& :=\av{\Lambda,i \tfrac{dU}{ds}\da U- \alpha H},
  \\
  \label{eq-LC}
  L_C &:= \alpha   \sum_j{\lambda_j}f^j(H), 
\end{align}
where $\angl{A,B}:=\Tr (\da AB)$.
The Hermitian operator
$\Lambda(t)$ and the real functions $\lambda$ and $\lambda_j(t)$ are Lagrange multipliers, while
the Hamiltonian $H(t)$ and the evolution operator $U(t)$ are dynamical variables.
The quantity $\log N$ represents the number of qubits. 
Also $U(T)$ is a dynamical variable, since the goal in the QBF problem is to obtain the time-optimal trajectories reaching the target $U_f$ only up to a fidelity $f$. 
The quantity $\alpha$ is the time cost. It plays the role of a "lapse" function \cite{mixed} connecting the physical time 
$t:=\int \alpha(s) ds$ to the parameter time $s$. 

The fidelity constraint is taken into account by the first term in the 
action $S$, and from variation with respect to $\lambda$ we obtain:
\begin{align}
|\av{U(T),U_f }|^2=(Nf)^2.  
\label{eq-fidelity}
\end{align}

Furthermore, variation of $L_S$ by $\Lambda$ gives the Schr\"odinger equation:
\begin{align}
  \label{eq-Sch}
  i\frac{dU}{dt}=HU, \quad\text{or}\quad 
  U(t)={\mathcal  T}e^{-i\int^t_0 Hdt}, 
\end{align}
where $\mathcal T$ is the time ordered product. 

Variation of $L_C$ by $\lambda_j$ leads to the constraints for $H$:
\begin{align}
  f_j(H)=0. 
\label{constraints}  
\end{align}
In particular, the finite energy condition reads: 
\begin{align}
f_0(H) :=\tfrac{1}{2}[\Tr(H^2)- N\omega^2]=0,
\label{eq-normH}
\end{align} 
where $\omega$ is a constant.

From the variation of $S$ with respect to $H$ we get:
\begin{align}
\Lambda =\lambda_0 H + \sum_{j\not = 0} \lambda_j \frac{\partial f_j(H)}{\partial H},
\label{eq-H}
\end{align} 
where we have used \Ref{eq-normH}.
From the variation of $S$ by $\alpha$, upon using \Ref{constraints} and \Ref{eq-H}, we obtain the normalization
condition:
\begin{align}
\Tr (H\Lambda)=N,
\label{eq-alpha}
\end{align}
while variation of $S$ with respect to $U(T)$ gives the following final boundary conditions for $\Lambda$:
\begin{align}
  \label{eq-Lbound}
  \Lambda(T)= -2\lambda~{\mathrm{Im}} [ \av{U(T), U_f} ~U(T)\da U_f ].
\end{align}

Finally, variation of $S$ by $U$, use of eq. \Ref{eq-H} and a simple algebra give the 
{\em quantum brachistochrone} equation: 
\begin{align}
  \label{eq-fund}
  i\frac{d\Lambda}{dt}= [H, \Lambda].
  \end{align}

The {\it quantum brachistochrone} together with the Schr\"odinger equation and the constraints define a two-boundary-value problem for the evolution of the unitary operator $U(t)$ and the Lagrange multiplier $\Lambda(t)$ with fixed initial $U(t=0)=1$ (where $1$ is the identity matrix) and final conditions $\Lambda(t=T)$ (eq. \Ref{eq-Lbound}), respectively.
For a given target gate $U_f$ and fidelity $f$, the optimal Hamiltonian $H$ and the optimal time duration $T$ can be found in the following way: a) solve the quantum brachistochrone \Ref{eq-fund} backwards in time subject to the constraints \Ref{constraints} and the final boundary condition \Ref{eq-Lbound} to obtain $H_{\mathrm{opt}}(t)$; b) integrate the Schr\"odinger equation \Ref{eq-Sch} forward in time with initial boundary condition $U(0)=1$ to get $U_{\mathrm{opt}}(t)$; c)
determine the integration constants in $H_{\mathrm{opt}}(t)$ by imposing the fidelity condition \Ref{eq-fidelity}.

\section{The QBF and an Ising Hamiltonian} 

We now explicitly solve the QBF problem for the physical system of three qubits
(labeled by  a superscript $a\in \{1, 2,  3\}$) arranged in a linear chain, interacting via an Ising Hamiltonian with 
interaction couplings $J_{12}, J_{23}$ and subject  to a local and controllable magnetic field $B_i(t)$ ($i=x, y, z$) acting 
on the intermediate qubit. 
More in details, we work with the Ising Hamiltonian:
\begin{align}
 H(t) :=J_{12}\sigma_z^{1}\sigma_z^{2} +J_{23}\sigma_z^{2}\sigma_z^{3} +\vec{B}(t)\cdot \vec{\sigma}^{2},
\label{ising}
\end{align}
where we have defined, e.g., $\sigma_i^1\sigma_j^2:=\sigma_i\otimes \sigma_j\otimes 1$,
the $\sigma_i$ are Pauli operators and $N=8$.
We simplify the notation introducing the ratio $K:=J_{23}/J_{12}$ and rescaling the time as $\tau:=J_{12}t$, 
the energy as $\hat\omega :=\omega/J_{12}$, the magnetic field as $\hat B(\tau):=B(t)/J_{12}$.
% and the Lagrange multiplier as $\hat \lambda_0:=J_{12}\lambda_0$. 

Our goal is to time-optimally generate the symmetric entangler gate $U^s_{13}$  acting between the indirectly coupled qubits 1 and 3 (see eq. (4) in \cite{khaneja2}), i.e. the target is: 
\begin{align}
U_f =U^s_{13}: = e^{-i\frac{\pi}{4}(\sigma^1_z\sigma^3_z+\sigma^1_z+\sigma^3_z)}.
\label{ufus}
\end{align}

The form \Ref{ising} of the physical Hamiltonian is guaranteed by the operator \Ref{eq-H}: 
\begin{eqnarray}
 \Lambda(t)&=\lambda_0 H + \sum_{i, j, k}\lambda_{ijk}\sigma_i^1 \sigma_j^2\sigma_k^3
 + \sum_i [\eta_i\sigma_i^1 + \xi_i\sigma_i^3] 
 \nonumber \\
 & + \sum_{i, j}[ \mu_{ij} \sigma_i^1\sigma_j^3
  +  \nu_{ij}\sigma_i^1\sigma_j^2
  + \rho_{ij} \sigma_i^2\sigma_j^3],
\label{fheisenberg}
\end{eqnarray}
where $\lambda_{ijk}(t), \mu_{ij}(t), \nu_{ij}(t), \rho_{ij}(t), \eta_i(t)$ and  $\xi_i(t)$ are Lagrange multipliers,
and the indices $\{i, j\}\in \{x, y, z\}$.
The finite energy condition \Ref{eq-normH} explicitly reads:
\begin{equation}
 \vec{\hat B}^2=\omega_K^2:=\hat \omega^2 -(1+K^2)=\mathrm{const},
 \label{normHk}
\end{equation}
and we must be sure that the available energy is enough so as to guarantee that $\omega_K^2 > 0$.

The relevant set of equations associated to the quantum brachistochrone \Ref{eq-fund} with $H$ and $\Lambda$ respectively given by \Ref{ising} and \Ref{fheisenberg} are presented as eqs. \Ref{f1}-\Ref{f5} in the Appendix A.
In this Appendix we also show that $\lambda_0$ and $\nu_{zz}+K\rho_{zz}$ are integrals of the motion.
Therefore, one can simplify the discussion by fixing the gauge freedom inherent to \Ref{eq-action} and choosing 
$\lambda_0=1$.
From this and the quantum brachistochrone equation, 
it is then immediate to find out that $\hat B_z=\mathrm{const}$ and therefore,
from the energy constraint \Ref{normHk}, that  $\hat B_x^2+\hat B_y^2 := \hat B_0^2=\mathrm{const}$.
Exploiting these integrals of the motion and after some lengthy but elementary algebra, we find that the general and non trivial solution of the quantum brachistochrone equation \Ref{eq-fund} is given by the Hamiltonian \Ref{ising} with the time-optimal magnetic field:
\begin{align}
  \vec{\hat B}_{\mathrm{opt}}(\tau)= \left ( \begin{array}{c}
 {\hat B_0}\cos \theta(\tau)\\
 {\hat B_0}\sin \theta(\tau)\\
   \hat B_z   
  \end{array}\right ),
  \label{bopt}
\end{align}
precessing around the $z$-axis with the frequency $\hat \Omega$, where $\theta(\tau):=\hat \Omega \tau +\theta_0$ and $\hat\Omega :=\Omega/J_{12}$ and $\theta_0$ are integration constants.
For later use, it turns convenient to use the energy constraint \Ref{normHk} and to define (for $\phi\in [0, 2\pi]$):
\begin{align}
\hat B_0&:=\omega_K \cos\phi;~~~~~~~\hat B_z:= \omega_K \sin \phi.
\label{Bphi}
\end{align}

The next step is to integrate the Schr\"odinger equation \Ref{eq-Sch} and to get the time-optimal evolution operator.
The details of the calculation are shown in the Appendix B, and the result can be summarized as:
\begin{align}
U_{\mathrm{\mathrm{opt}}}(\tau)=[a^{13}(\tau)-i\vec{b}^{13}(\tau)\cdot \vec{\sigma}^2],
\label{uopt}
\end{align}
where the operators $a^{13}(\tau)$ and $\vec{b}^{13}(\tau)$ act in the Hilbert space of qubits 1 and 3 and 
are defined in eqs. \Ref{ab} of the Appendix B.

Finally, one has to impose the boundary conditions \Ref{eq-Lbound} on $\Lambda$ at the final time $\tau_\ast :=J_{12}T$.
Again, after some lengthy but simple algebra (for the explicit details the reader is referred to the Appendix C), it turns out that 
in the case of our 3-qubit model the relevant conditions can be summarized into the following equations:
\begin{align}
\hat\Omega\tau_\ast=2m\pi,
 \label{omegaopt}
\end{align}
with $m$ a non zero integer \footnote{If $m=0$, either $T=0$ or $\Omega=0$, which are both trivial cases.}
and 
\begin{align}
|M(\tau_\ast)|&=4f,
\label{mopt}
\\
P(\tau_\ast)&=R(\tau_\ast),
\label{popt}
\\
Q(\tau_\ast)&=\frac{\hat\Omega}{2}P(\tau_\ast),
 \label{qopt}
\end{align}
where the functions $M(\tau_\ast), P(\tau_\ast), Q(\tau_\ast)$ and $R(\tau_\ast)$ are defined in eqs. \Ref{M}-\Ref{R}
of the Appendix  C.
In other words, the QBF problem can be reduced to find the unknown integrals of the motion
$\hat B_z, \hat B_0, \hat\Omega , \theta_0$ and $\tau_\ast$ from the constraints \Ref{omegaopt}-\Ref{qopt} 
(via eqs. \Ref{omegak} and \Ref{bi}, see the Appendix C).

\section{Perfect matching: $f=1$ }

We first consider the case of perfect matching between the time-optimal evolution operator 
$U_{\mathrm{opt}}(\tau_\ast)$ and the target $U_f=U_s^{13}$.
In other words we assume that the fidelity is the maximum achievable, i.e. $f=1$.
In this case, it is immediate to see from eqs. \Ref{mopt}, \Ref{sc}-\Ref{bi} and  \Ref{M} that the only possibility is
to have:
\begin{align}
\omega_i\tau_\ast=\pi n_i~;~~~~~~i=1, 2, 3, 4,
 \label{f=1}
\end{align}
where $n_1, n_2, n_3, n_4$ are positive integers such that $n_2=n_1+2q-1$, $n_3=n_1+2s-1$ and $n_4=n_1+2r-1$, with 
$q, r, s$ arbitrary integers. 
Then, from eqs. \Ref{f=1} and \Ref{sc} we find that all $s_i$s are zero, and from eqs. \Ref{P}-\Ref{R} we see that also 
$P(\tau_\ast)=Q(\tau_\ast)=R(\tau_\ast)=0$, while $M(\tau_\ast)= 4(-1)^{n_1}$.
Therefore, eqs. \Ref{mopt}-\Ref{qopt} are automatically satisfied, and one is left with the conditions \Ref{omegaopt} and
\Ref{f=1}, to be satisfied via eqs. \Ref{omegak} and \Ref{bi}. 
This is essentially the same result which we had already found in our previous work \cite{ising} for the case of equal couplings $K=1$ (look at eqs. (43) and (63) in \cite{ising}).
In particular, following exactly the same methods of \cite{ising}, one first multiplies eqs. \Ref{omegak} by $\tau_\ast$, 
and then, using eqs. \Ref{f=1}, inverts the former to find (with the help of \Ref{Bphi}) the equivalent of eqs. (47)-(50) 
of \cite{ising}, i.e.:
\begin{align}
\tau_\ast&=\pi\sqrt{\frac{f_-}{8K}},
\label{tau}
\\
\hat \Omega &= 2m\sqrt{\frac{8K}{f_-}},
\label{OOmega}
\\
(\hat B_z)^2&=\frac{8K}{f_-}\left (m- \sqrt{\frac{g_+g_-}{8f_-}} \right )^2,
\label{Bz}
\\
(\hat B_0)^2&=\frac{K}{f_-}\left [2\Delta f-\frac{(\Delta g)^2}{g_+g_-}f_- -\frac{g_+g_-}{f_-} \right ],
\label{B0}
\end{align}
where we have introduced:
 \begin{align}
 f_\pm &:= (n_1^2+n_4^2)\pm (n_2^2+n_3^2)
 \\
 g_\pm &:= (n_1^2-n_4^2)\pm (n_2^2-n_3^2)
 \label{fg}
\end{align}
and their differences $\Delta f :=f_+-f_-, ~\Delta g:= g_+-g_-$.
Moreover, the energy and the Ising interaction couplings are subject to the constraint:
\begin{align}
\omega_K^2=\frac{2K}{f_-}\left [f_+ +4m\left (m-\sqrt{\frac{g_+g_-}{2f_-}}\right)\right ] -(1+K^2),
 \label{omk}
\end{align}
the values of Ising couplings are also constrained by the relation $K=g_-/g_+$
\footnote{In fact, when the couplings are equal, then $g_+=g_-$ and coincide with $f_0$ of eq. (46) in \cite{ising}.
Then from \Ref{fg} $n_2=n_3$ so that also the $f_\pm$ coincide with eq. (45) of \cite{ising}.}, 
and one has to guarantee that $(\hat B_0)^2\geq 0$ and that $\omega_K^2  > 0$ for the existence
of the solutions \footnote{This, in particular, requires that $\mathrm{sign}(f_-)=\mathrm{sign}(K)=\mathrm{sign}(g_+g_-)$.}.

Again, after a long but straightforward analysis (see \cite{ising}), one can show that the perfect ($f=1$) time-optimal 
generation of the gate $U^s_{13}$ is possible for the values of $f_-=3$ and $n_1$ even, and the solution coincides 
with that given in eqs. (51)-(54) of \cite{ising} for the case of equal couplings.
In particular, for $n_1=2$ and $n_2=n_3=n_4=1$ one gets: 
\begin{align}
\tau_{\ast \mathrm{opt}}&= \pi\sqrt{\frac{3}{8}},
\label{tauopt}
\\
\hat\Omega_{\mathrm{opt}} &=2m\sqrt{\frac{8}{3}},
\label{OOmegaopt}
\\
|\hat B_z|_{\mathrm{opt}}&=\biggl |\sqrt{\frac{8}{3}}m-1\biggr |,
\label{Bzopt}
\\
|\hat B_0|_{\mathrm{opt}}&=\sqrt{\frac{5}{3}},
\label{B0opt}
\end{align}
and 
\begin{align}
\omega_1^2=\frac{8}{3}\left [1-\sqrt{\frac{3}{2}}~m+m^2\right].
\label{omkopt}
\end{align}
We stress that the duration time \Ref{tauopt} is shorter than the numerical result proposed in the earlier literature \cite{khaneja2} for the same trilinear Ising Hamiltonian and target as ours but for local controls which can selectively and
{\it instantaneously} address the single qubits of the chain. In other words, as already cautioned by their authors,
the quantum control procedure discussed in the paper \cite{khaneja2} is not time optimal (although it performs 
better than other standard procedures).  

For unequal couplings, i.e. $K\not =1$, we could not find solutions of the equations \Ref{tau}-\Ref{omk} which satisfy the conditions $(\hat B_0)^2\geq 0$, $~\omega_K^2>0$ and $K=g_-/g_+$ for $f_-$ smaller than $7$.
The reason of this unexpected result for the $K\not =1$ case can be found by taking into consideration the more general
ansatz of almost perfect matching between $U_{\mathrm{opt}}(\tau_\ast)$ and the target $U_f$, i.e. the ansatz of a fidelity $f\lessapprox 1$. This is the topic of the next Section.  

\section{Almost perfect matching: $f\lessapprox1$ }

We are now interested in finding the solution to the QBF problem in the case when the required final fidelity is not exactly one, but we allow for some error.
In other words, we want to achieve a final fidelity: 
\begin{align}
f=1-\frac{\epsilon^2}{8},
 \label{fepsilon}
\end{align}
where $|\epsilon| \ll 1$.
By looking at the final boundary condition \Ref{mopt} and taking into account \Ref{fepsilon}, it is immediate to verify that the QBF problem can be solved by generalizing the ansatz \Ref{f=1} to:
\begin{align}
\omega_i\tau_\ast =\pi n_i +\sigma_i~;~~~~i=1, 2, 3, 4,
 \label{fnot1}
\end{align}
where the real variables $\sigma_i\ll 1$, while the integers $n_i$ satisfy the same relations as those written below 
eq. \Ref{f=1}.

Then, substituting \Ref{fnot1} into eqs. \Ref{mopt}-\Ref{qopt}, we obtain the following constraints for the $\sigma_i$s: 
\begin{align}
[b_1-b_4]\biggl [\frac{b_1\sigma_1}{n_1}-\frac{b_4\sigma_4}{n_4}\biggr ]=[b_3-b_2]\left [\frac{b_2\sigma_2}{n_2}-\frac{b_3\sigma_3}{n_3} \right ],
\label{13}
\end{align}
\begin{align}
(b_1-\sin\phi)\frac{\sigma_1}{n_1}+(b_2-\sin\phi)\frac{\sigma_2}{n_2}&-(b_3-\sin\phi)\frac{\sigma_3}{n_3}
\nonumber \\
&=(b_4-\sin\phi)\frac{\sigma_4}{n_4},
\label{15}
\end{align}
and
\begin{align}
\sigma_1^2+\sigma_2^2+\sigma_3^2+\sigma_4^2=\epsilon^2.
 \label{22}
\end{align}
In principle, the procedure is now similar to that of the previous Section.
In other words, one should invert eqs. \Ref{omegak} in order to express the unknown constants of the motion 
$\hat B_0, \hat B_z, \hat \Omega$ and $\tau_\ast$ in terms of the integers $n_i$ and the variables $\sigma_i$,
and then determine the former (constants) by minimizing $\tau_\ast$ over the latter (integers).
However, as it is shown in the Appendix D, only for values of the couplings ratio $|K|$ close to one it is possible 
to find time-optimal solutions.
For pedagogical purposes, without losing generality we present the prototypical ansatz of positive $K$, i.e.
$K=1+\delta K$, with $|\delta K|\ll 1$.
The case of negative $K$ can be dealt with along similar lines \footnote{In this case, for the reality of the duration time,
eq. \Ref{tauopt}, one has to work with $f_-$ negative.}.

Then, performing a perturbative expansion in $\delta K$, after some simple and tedious algebra (see Appendix D) we
obtain a self-consistent set of values for the variables $\sigma_i$ (eqs. \Ref{sigma234}-\Ref{sigma1}
of the Appendix D) which, for a given $K$ close to one, leads to the following time-optimal constants of the
unitary evolution: 
\begin{align}
\tau_{\ast \mathrm{opt}} &\simeq \pi\sqrt{\frac{3}{8}}\biggl (1-\frac{\delta_{\epsilon K}}{2}\biggr),
 \label{opttau}
\\
\hat \Omega_{\mathrm{opt}} &\simeq 2\sqrt{\frac{8}{3}}\biggl (1+\frac{\delta_{\epsilon K}}{2}\biggr ),
\label{optOmega}
\\
\hat B_{z, \mathrm{opt}} &\simeq \left (1+\sqrt{\frac{8}{3}}\right )\biggl [1+\frac{(\sqrt{2}+\sqrt{3})}{(2\sqrt{2}+\sqrt{3})}
\delta_{\epsilon K}\biggr ],
\label{optBz}
\\
|\hat B_0|_{\mathrm{opt}} &\simeq \sqrt{\frac{5}{3}}\biggl \{1+\frac{1}{5}\biggl [\delta K
\nonumber \\
&+\frac{2(12\sqrt{6}+31)}{\pi}\sqrt{\frac{\Delta_{\epsilon K}}{27+4\sqrt{6}}}\biggr ]\biggr \},
\label{optB0}
\end{align}
where we have defined:
\begin{align}
\Delta_{\epsilon K}&:=\epsilon^2-\frac{9\pi^2}{32}(\delta K)^2,
\label{}
\\
\delta_{\epsilon K}&:= \delta K -\frac{2(1+2\sqrt{6})}{3\pi}\sqrt{\frac{\Delta_{\epsilon K}}{27+4\sqrt{6}}}.
\label{}
\end{align}
We point our that the optimal time duration \Ref{opttau} is approximately a function of $K^{-1/2}$, i.e. it decreases 
for increasing $K$.
The optimal values of the integers are again $n_1=2$, $n_2=n_3=n_4=1$, for which $f_-=3$, and in this case also the 
integer $m$ can be optimized (see Appendix D) as $m=1$.
From the above equations, it is also clear that, in order for the time-optimal evolution to exist, i.e. for the values of
$\tau_\ast, \hat B_z, \hat B_0$ and $\hat \Omega$ to to be real, the ratio of the couplings in the Ising Hamiltonian 
must satisfy the condition $\Delta_{\epsilon K}\geq 0$ or, in terms of the fidelity, using \Ref{fepsilon}:
\begin{align}
|\delta K|\leq \frac{16}{3\pi}\sqrt{1-f}.
 \label{Kepsilon}
\end{align}
This means that, the better we want to approximate the unitary target, i.e. the closer the fidelity should be to one,
the smaller the deviation from one of the ratio between the Ising couplings that we can choose so that the perturbative expansion and the time-optimal evolution laws are self-consistent. 
Alternatively, if we have an experiment where $K$ is given, we can read formula \Ref{Kepsilon} as predicting
that the distance of $K$ from the unit value determines the maximal
fidelity that we can hope to achieve (and impose in the QBF formalism).
Furthermore, we have to choose an energy which satisfies the following constraint:
\begin{align}
\hat \omega_K^2&\simeq \frac{16}{3}\left (1+\sqrt{\frac{3}{8}}\right )
\biggl \{1+\frac{(4\sqrt{2}+3\sqrt{3})}{2(2\sqrt{2}+\sqrt{3})}
\nonumber \\
&\cdot\biggl [\delta K +\frac{(\sqrt{6}+29)}{15\pi}\sqrt{\frac{\Delta_{\epsilon K}}{27+4\sqrt{6}}}   \biggr ]\biggr \}.
 \label{optomegak}
\end{align}

We give a few examples of concrete situations where our formalism can operate and give the shortest time-duration of
the quantum control in order to achieve the gate $U_s^{13}$ with fixed energy and fidelity.
For instance, in the context of NMR \cite{jones}, we can think of an 
experiment where the Ising chain is the
$H-N-H$ part of the molecule of ethanamide \cite{nimbalkar}, for which the Ising couplings are $J_{12}=J_{23}
\simeq 88.05$ Hz, or the $F-F-F$ part of the trifluoroaniline molecule \cite{mangold}, for which the Ising 
couplings are $J_{12}=J_{23} \simeq 20$ Hz. For these molecules, $K=1$ and the time optimal duration is given
by eq. \Ref{tauopt} of the previous Section.
On the other hand, we can imagine an NMR experiment based on the $H-H-H$ chain in the molecule of the 
%2,3 dibromopropanoic acid \cite{linden}, 
1-chloro-2-nitro-benzene \cite{cory2}, where the Ising couplings are $J_{12}\simeq 8$ Hz and $J_{23} \simeq 7$ Hz,  
or an experiment based on the $H-H-P$ chain in the molecule of 2-chloroethenylphoshonic acid \cite{cummins},
 with $J_{12}\simeq 9.1$ Hz, $J_{23}\simeq 11.3$ Hz.
%$H-C-F$ chain from the molecule of diethylfluoromalonate \cite{peng}, which has  $J_{12}\simeq 161.3$ Hz and 
%$J_{23}\simeq -192.2$ Hz.
The former has a ratio $K\simeq 0.875$, and the optimal fidelity reachable according to \Ref{Kepsilon} is 
$f_{\mathrm{max}}\simeq 0.995$, within a optimal time duration given by eq. \Ref{opttau}.
The latter has a ratio $K\simeq 0.805$, and the optimal fidelity reachable according to \Ref{Kepsilon} is 
$f_{\mathrm{max}}\simeq 0.987$.
%Other interesting experimental tests of the QBT formalism in the NMR context might come from the use of  of the $F-F-F$ %chain in the iodotrifluoroethylene molecule \cite{du}, with $J_{12}\simeq 64.2$ Hz, $J_{23}\simeq 51.3$ Hz and a ratio $K%\simeq 0.799$.
We notice that even taking the popular NMR model of the $C-C-C$ chain the alanine molecule \cite{cory},
with $J_{12}\simeq 54$ Hz, $J_{23}\simeq 35$ Hz and a relatively large $\delta K\simeq 0.35$,  
the maximal achievable fidelity is still $f_{\mathrm{max}}\simeq 0.957$.
It would be also interesting to apply the QBT formalism to experiments involving, e.g., supeconducting qubits.

\section{Changing coordinates: $\mathrm{CNOT}(1, 3)$}

An analysis similar to that performed in the last Section can be done in the following two situations.
On the one hand, the physical Hamiltonian \Ref{ising} may be available in an experiment performed within the standard computational basis (the Hilbert space spanned by the states $\{\ket{0}, \ket{1}\}\otimes \{\ket{0}, \ket{1}\}\otimes \{\ket{0}, \ket{1}\}$), but from the point of view of another lab the Hilbert space is seen as being spanned by the states $\{\ket{0}, \ket{1}\}\otimes \{\ket{0}, \ket{1}\}\otimes\{\ket{+}, \ket{-}\}$, where the basis for qubit 3 is $\{\ket{+}, \ket{-}\}$ (with $\ket{+}:=W\ket{0}=(\ket{0}+\ket{1})/\sqrt{2}; ~\ket{-}:=W\ket{1}=(\ket{0}-\ket{1})/\sqrt{2}$ ($W$ is the Walsh-Hadamard transform). Thus, the second lab sees an effective, rotated Hamiltonian given by:
 \begin{align}
 H^\prime(\tau)=\sigma_z^{1}\sigma_z^{2} +K\sigma_z^{2}\sigma_x^{3} +\vec{\hat B}(\tau)\cdot \vec{\sigma}^{2}.
\label{ising2}
\end{align}
Alternatively, one may simply assume that the Hamiltonian \Ref{ising2} is available in an experiment performed within the 
standard computational basis for all the qubits.  
The Hamiltonians \Ref{ising} and \Ref{ising2} are related by:
\begin{align}
H^\prime(\tau)=V^3H(\tau)V^3,
\label{rotatedh}
\end{align}
where $V^3:=1\otimes 1\otimes W$.
Now, the goal is to time-optimally synthesize with fixed fidelity the gate $\mathrm{CNOT}(1, 3)$:
\begin{align}
U_f:=  CNOT(1, 3) = e^{-i\frac{\pi}{4}(1+ \sigma^1_z\sigma^3_x-\sigma^1_z-\sigma^3_x)}.
\label{uf1}
\end{align}

As explained in our previous work \cite{ising}, an analysis parallel to that of the previous Sections can be made.
The relevant quantum brachistochrone equations are still given by \Ref{f1}-\Ref{f5}, with the only replacement
$\rho_{iz}\rightarrow \rho_{ix}$.
Then the time-optimal evolution operator becomes: 
\begin{align}
U'_{\mathrm{opt}}(\tau)&=V^3U_{\mathrm{opt}}(\tau)V^3,
\end{align}
where $U_{\mathrm{opt}}(t)$ is given by \Ref{uopt}. 

Also the new target $\mathrm{CNOT}(1, 3)$ can be diagonalized (in the 1, 3 qubit subspace) as:
\begin{align}
U_f=V^3{U'}_D^{13} V^3=V^3\mathrm{Diag}(1, 1, 1, -1)V^3,
\label{uf2}
\end{align}
where the operator ${U'}_D^{13}$ acts in the 1,3 qubit subspace.
Then, following the same methods of the previous Section, it is easy to show that the only relevant changes are in a flip of the signs in front of the terms $s_1$ and $s_4$ in eqs. \Ref{M}, \Ref{P} and \Ref{R}.
Furthermore, in formula \Ref{omegak} we have choose the new positive integers $n_i^\prime ~, i=1, 2, 3, 4$ with 
$n^\prime_2 =n^\prime_ 1+2q^\prime $, $n^\prime_3 =n^\prime_ 1+2s^\prime$ and $n^\prime_4=n^\prime_1 +2r^\prime +1$, where $q^\prime, s^\prime, r^\prime$ are arbitrary integers.
Considerations similar to those made in the previous Section finally lead to the expressions
\Ref{tau}-\Ref{B0}, and to the minimization of the evolution time $\tau_\ast$ for $f_-=3$ and $n_1$ odd.
In particular, for $n_1=n_2=n_3=1$ and $n_4=2$, the solution to the QBF problem with perfect matching ($f=1$) between 
$U_{\mathrm{opt}}(\tau_\ast)$ and the $\mathrm{CNOT}(1, 3)$ target is given again, for equal couplings $K=1$, by 
eqs. \Ref{tauopt}-\Ref{omkopt} (cf. eqs. (51)-(55) of \cite{ising}).
For different couplings, the perturbative expansion in $\delta K$ leads instead to the time-optimal solution
\Ref{opttau}-\Ref{optomegak}, except for a global change of sign for $\hat\Omega$ and $\hat B_z$ (see Appendix D).

\section{Discussion}

We have presented the general theoretical framework for determining the time-optimal way of generating a target unitary quantum gate when the physical Hamiltonian is subject to certain constraints (at least a fixed energy available) 
and when some errors on the target are tolerated, i.e. the fidelity is assumed to be fixed and smaller than one. 
We have then applied the formalism, which for brevity we called QBT, to the model study of a trilinear Ising chain 
with unequal Ising interaction couplings and where the intermediate qubit is subject to coherent control
via a local, but in principle time consuming, magnetic field.
For the case of a fixed fidelity equal to one (i.e. a perfect matching with the target), we explicitly gave analytical formulae
of the time-optimal control laws and of the minimal time required to realize the entangling gate $U_s^{13}$ between the indirectly coupled qubits of the chain when the Ising couplings are equal.
The analytical expression for the minimal time found is shorter than the numerical expression found (using 
standard geometrical quantum control methods for the same trilinear Ising Hamiltonian plus instantaneous 
individual control of the qubits) in \cite{khaneja2}.
For the case of a fixed fidelity smaller than one, we performed an perturbative expansion around a ratio between the
Ising couplings equal to one, and we gave analytical results for the time-optimal control laws and for the minimal
time required to achieve $U_s^{13}$ for unequal couplings. 
Changing the computational basis for one of the qubits at the end of the chain, similar methods can be used
to analytically express the time-optimal control laws for generating the $\mathrm{CNOT}$ gate between the 
indirectly coupled qubits. 
All of the above results are new, analytical and constitute an improvement with respect to the present literature 
for the values of the minimal time required to generate certain unitary quantum gates between indirectly coupled qubits. 
We also proposed a series of experiments within the NMR paradigm to test our theoretical predictions.
The main advantage inherent to the use of the QBF formalism appears to lay in the fact that (at least for these
simple low dimensional models) it naturally allows for local quantum operations which in principle may have a non 
zero time cost, while  the standard geometrical methods usually assume that local operations should be done instantaneously.
In some sense, our ansatz is slightly more restricted than that of \cite{khaneja2}, since we also impose 
the finite energy condition eq. \Ref{eq-normH}.
This, in turns, determines the appearance of the constraint \Ref{omkopt} (or \Ref{optomegak}), which 
relates the allowed values of the Ising couplings to the energy available in the experiment. 
As we already pointed out in \cite{ising}, we conjecture that this feature is the consequence 
of working with equality constraints on the Hamiltonian in the action principle, and we expect that
it should be mitigated when such an assumption is relaxed and one considers inequality constraints 
as well (for example, an energy available bounded from above).
We conjecture that also the difficulty in finding explicit results in the case of arbitrary $K$ far from one
stems from the same technical point.
We plan to extend our methods to the study of chains with more qubits and more complex topologies,
to extend the allowed local controls to all qubits and to consider other quantum gates. 
Also in progress is some work related to the speed of quantum information transmission
(e.g. transfer of coherence) along spin chains, and to the role of entanglement during the QBF evolution.

\section{Appendix}

\subsection{QB and Integrals of Motion}

Comparing the coefficients of the generators of ${\alg su}(8)$ on both sides of \Ref{eq-fund}, with $\Lambda$ given by 
\Ref{fheisenberg} and $H$ given by \Ref{ising}, we find that the relevant quantum brachistochrone equations are:
\begin{align}
({\hat\lambda_0\hat B}_x)^\cdot&=-2(\nu_{zy} +K\rho_{yz}),
\label{f1}\\
({\hat\lambda_0\hat B}_y)^\cdot&=2 (\nu_{zx} +K\rho_{xz}),
\label{f2}\\
({\hat\lambda_0\hat B}_z)^\cdot&=0,
\label{f3}\\
\dot{\nu}_{zz}&=2(B_x\nu_{zy} -B_y\nu_{zx}),
\label{f4}\\
\dot{\rho}_{zz}&=2(B_x\rho_{yz} -B_y\rho_{xz}),
\label{f5}
\end{align}
where $\hat \lambda_0:=J_{12}\lambda_0$ is a rescaled Lagrange multiplier. 
First of all, using the energy constraint equation \Ref{eq-normH} we note that 
$({\vec{\hat B}})^2=\hat B_x^2 +\hat B_y^2+\hat B_z^2=\mathrm{const}$.
Moreover, from \Ref{eq-H} , the quantum brachistochrone \Ref{eq-fund} and \Ref{fheisenberg}, we can compute 
$0=\Tr [\dot\Lambda H]=8[\hat\omega^2\dot{\hat\lambda}_0+(\dot \nu_{zz}+K\dot \rho_{zz})]$.
Then, taking the time derivative of $({\vec{\hat B}})^2$, using eqs. \Ref{f1}-\Ref{f5} and the latter result, a simple algebra
shows that $\hat\lambda_0=d_1$ and $\nu_{zz}+K\rho_{zz}=d_2$, where $d_1, d_2$ are constants.
Finally, imposing the normalization condition \Ref{eq-alpha}, one finds that $d_1\hat\omega^2+d_2=1$.

\subsection{Time-optimal Evolution Operator}

We now proceed to integrate the Schr\"odinger eq. \Ref{eq-Sch} for $U_{\mathrm{opt}}(\tau)$, given that $H_{\mathrm{opt}}(\tau)$ is expressed via eqs. \Ref{ising} and \Ref{bopt}.
For this purpose, we exploit the following well known rotation formula for the Pauli matrices:
\begin{align}
e^{-i\frac{\theta(\tau)}{2}\sigma_z}\sigma_x e^{i\frac{\theta(\tau)}{2}\sigma_z}=\cos \theta(\tau) \sigma_x +\sin \theta(\tau)\sigma_y,
\end{align}
and we rewrite the time-optimal Hamiltonian as:
\begin{align}
H_{\mathrm{opt}}(\tau)=e^{-i\frac{\theta(\tau)}{2}\sigma_z^2} H_0~ e^{i\frac{\theta(\tau)}{2}\sigma_z^2},
\label{hopt}
\end{align}
where we have introduced the constant operator: 
\begin{eqnarray}
H_0:=\hat B_0\sigma_x^2 +(\sigma_z^1 +K\sigma_z^3+\hat B_z)\sigma_z^2.
\end{eqnarray}
Furthermore, defining the transformed unitary operator:
\begin{align}
\tilde{U}(\tau):=e^{i\frac{\theta(\tau)}{2}\sigma_z^2}U(\tau),
\label{tildeU}
\end{align}
we easily check that, since $U(\tau)$ should obey the Schr\"odinger equation \Ref{eq-Sch}, $\tilde{U}(\tau)$ should also satisfy:
\begin{align}
i\frac{d\tilde{U}}{d\tau}=\tilde{H}\tilde{U}
\label{schrnew}
\end{align}
with the time-independent Hamiltonian:
\begin{align}
\tilde{H}:=H_0-\frac{\hat \Omega}{2}\sigma_z^2=\mathrm{const}.
\end{align}
We note that the constant Hamiltonian $\tilde{H}$ is diagonal in the 1,3 qubit subspace, i.e.,
\begin{align}
\tilde{H}=\hat B_0\sigma_x^2 + B_D^{13} \sigma_z^2=\mathrm{const},
\label{tildeH}
\end{align}
where we have introduced the operator $B_D^{13}$ acting in the 1,3 qubit subspace:
\begin{align}
B_D^{13}&:=\omega_K\mathrm{Diag}[b_1, b_2, b_3, b_4],
\label{bd}
\end{align}
which depends upon the constants:
\begin{align}
b_i&:=\sin\phi +\frac{1}{\omega_K}\biggl [(\delta_{i1}-\delta_{i4})(1+K)
\nonumber 
\\
&+(\delta_{i2}-\delta_{i3})(1-K)- \frac{\hat\Omega}{2}\biggr ],
\label{bi}
\end{align}
where $i=1,2,3,4$ and $\delta_{ij}$ is the Kronecker symbol.

Then, solving eq. \Ref{schrnew} together with eq. \Ref{tildeH} for $\tilde{U}(\tau)$ and finally inverting \Ref{tildeU}, it is easy to check that the time-optimal unitary operator $U_{\mathrm{opt}}(\tau)$ evolves as:
\begin{align}
U_{\mathrm{opt}}(\tau)=e^{-i\frac{\theta(\tau)}{2}\sigma_z^2}e^{-i\tilde{H}\tau}e^{i\frac{\theta_0}{2}\sigma_z^2}.
\label{uopt1}
\end{align}
In particular, the exponential of the constant Hamiltonian appearing on the right hand side of eq. \Ref{uopt1}
is also diagonal (in the 1, 3 qubit subspace) and can be expanded as:
\begin{align}
e^{-i\tilde{H}\tau}=C_D^{13}(\tau)-iS_D^{13}(\tau)\tilde{H},
\label{exptildeH}
\end{align}
where we have introduced the diagonal operators:
\begin{align}
S_D^{13}(\tau)&:=\mathrm{Diag}[s_1(\tau), s_2(\tau), s_3(\tau), s_4(\tau)],
\label{}\\
C_D^{13}(\tau)&:=\mathrm{Diag}[c_1(\tau), c_2(\tau), c_3(\tau), c_4(\tau)],
\label{scd}
\end{align}
which depend upon the functions of $\tau$: 
\begin{eqnarray}
s_i(\tau):=\frac{\sin (\omega_i \tau)}{\omega_i};~~~~~~~c_i(\tau):=\cos (\omega_i \tau),
\label{sc}
\end{eqnarray}
and the constants:
\begin{align}
\omega_i & :=\omega_K\sqrt{\cos^2\phi +b^2_i}.
\label{omegak}
\end{align}
Thus, inserting eq. \Ref{exptildeH} into eq. \Ref{uopt1}, one obtains the more explicit expression \Ref{uopt}
for the time-optimal evolution of the unitary operator, where we have defined:
\begin{align}
a^{13}(\tau)&=\cos\left [\frac{\hat\Omega \tau}{2}\right ]C_D^{13}(\tau)-\sin\left [\frac{\hat\Omega \tau}{2}\right ]B_D^{13}S_D^{13}(\tau),
\\
b^{13}_ x(\tau)&=B_0S_D^{13}(\tau)\cos\left [\frac{\hat\Omega \tau}{2} +\theta_0\right ],
\\
b^{13}_y(\tau)&=B_0S_D^{13}(\tau)\sin\left [\frac{\hat\Omega \tau}{2} +\theta_0\right ],
\\
b^{13}_z(\tau)&=\sin\left [\frac{\hat\Omega \tau}{2}\right ]C_D^{13}(\tau)+\cos\left [\frac{\hat\Omega \tau}{2}\right ]B_D^{13}S_D^{13}(\tau).
\label{ab}
\end{align}
The expression \Ref{uopt} still depends upon the integration constants 
$\hat B_0, \hat B_z$, $\hat \Omega$ and $\theta_0$ and the optimal duration time $\tau_\ast$ of the unitary evolution.
These parameters are determined by imposing the final fidelity constraint \Ref{eq-fidelity} and the final boundary condition \Ref{eq-Lbound}.

\subsection{Final Boundary Conditions on $\Lambda$}

The boundary conditions for $\Lambda$ at the final time $\tau_\ast :=J_{12}T$ in our 3-qubit model explicitly read:
\begin{align}
\Lambda(\tau_\ast)=2\lambda ~U_D^{13}~ \vec{b}^{13}(\tau_\ast)\cdot\vec{\sigma}^2~\Tr[U_{\mathrm{opt}}(\tau_\ast)\da U_f].
 \label{eq-Lambdabound1}
\end{align}
Let us now compute these conditions.
First, we note that our final target \Ref{ufus} is diagonal in the 1, 3 qubit subspace and can be written as:
\begin{align}
U_f =e^{i\frac{\pi}{4}}U_D^{13}, ~~~~U_D^{13}:= \mathrm{Diag}(-1, 1, 1, 1). 
\label{ufus1}
\end{align}
Then, using \Ref{uopt} and \Ref{ufus1}, we can write:
\begin{align}
 \Tr [U_{\mathrm{opt}}(\tau_\ast )U_f]=2\biggl(\cos\left [\frac{\hat\Omega \tau_\ast}{2}\right ]M
 - \sin\left [\frac{\hat\Omega \tau_\ast}{2}\right ]N\biggr ),
\label{traceutuf}
\end{align}
where we have defined:
\begin{align}
M(\tau_\ast)&:=\Tr [C_D^{13}U_D^{13}] =c_1+c_2+c_3-c_4,
\label{M}
\\
N(\tau_\ast)&:=\Tr [B_D^{13}S_D^{13}U_D^{13}]=b_1s_1+b_2s_2+b_3s_3-b_4s_4.
\label{N}
\end{align}

Now, inserting \Ref{traceutuf} into \Ref{eq-Lambdabound1}, from the $\sigma^2_x, \sigma^2_y, \sigma^2_z$ terms of $\Lambda(\tau_\ast)$, we obtain the final boundary conditions: 
\begin{align}
\cos \theta(\tau_\ast)&=\frac{\lambda P(\tau_\ast)}{2\omega_K}\cos\left [\frac{\hat\Omega \tau_\ast}{2}+\theta_0\right ]
\Tr [U_{\mathrm{opt}}(\tau_\ast )\da U_f],
\label{lambdaBx}
\\
\sin \theta(\tau_\ast)&=\frac{\lambda P(\tau_\ast)}{2\omega_K}\sin\left [\frac{\hat\Omega \tau_\ast}{2}+\theta_0\right ]
\Tr [U_{\mathrm{opt}}(\tau_\ast )\da U_f],
\label{lambdaBy}
\\
\hat B_z&=\frac{\lambda}{2}\biggl(\cos\left [\frac{\hat\Omega \tau_\ast}{2}\right ]N
+\sin\left [\frac{\hat\Omega \tau_\ast}{2}\right ]M\biggr )
\nonumber \\
&\cdot \Tr [U_{\mathrm{opt}}(\tau_\ast )U_f],
 \label{lambdaBz}
\end{align}
where  
\begin{align}
P(\tau_\ast)&:=\Tr [S_D^{13}U_D^{13}]=s_1+s_2+s_3-s_4.
\label{P}
\end{align}

Upon using \Ref{traceutuf} into \Ref{lambdaBx} and \Ref{lambdaBy} we find that these can be nontrivially \footnote{The trivial solutions are $\hat B_0$ or $\hat B_z$ equal to zero.} satisfied only if:
\begin{align}
\tan\theta(\tau_\ast)=\tan\left[\frac{\theta(\tau_\ast)+\theta_0}{2}\right ],
 \label{}
\end{align}
whose solution is given by:
\begin{align}
\hat\Omega\tau_\ast=2m\pi ,
 \label{Omega}
\end{align}
where $m$ is a non zero integer \footnote{If $m=0$, either $T=0$ or $\Omega=0$, which are both trivial cases.}.
Using the latter formula we can compactly rewrite equations \Ref{lambdaBx} -\Ref{lambdaBz} as:
\begin{align}
\lambda M(\tau_\ast)P(\tau_\ast) &=\omega_K,
\label{lmn}
\\
\lambda M(\tau_\ast)N(\tau_\ast) &=\hat B_z.
 \label{lmp}
\end{align}
Moreover, from the fidelity eq. \Ref{eq-fidelity} and the definition: 
\begin{align}
\Tr[U_{\mathrm{opt}}(\tau_\ast)\da U_f]:=|\Tr[U_{\mathrm{opt}}(\tau_\ast)\da U_f]|e^{-i\chi}=8f ~e^{-i\chi},
 \label{traceutufdef}
\end{align}
and from \Ref{traceutuf} and \Ref{traceutufdef}, we obtain:
\begin{align}
e^{i\chi}&=(-1)^m\mathrm{sign}[M(\tau_\ast)]
 \label{m1modsign}
\end{align}
and eq. \Ref{mopt}.

Let us now impose the normalization condition \Ref{eq-alpha} at the final time $\tau_\ast$.
Multiplying \Ref{eq-Lbound} by $H(\tau_\ast)$ and using \Ref{eq-alpha}, we
can express the Lagrange multiplier $\lambda$ as:
\begin{align}
\lambda=-\{2~{\mathrm{Im}} (e^{i\chi}\Tr[U_{\mathrm{opt}}(\tau_\ast)\da U_f H_{\mathrm{opt}}(\tau_\ast)])\}^{-1}.
 \label{lambda}
\end{align}
After some lengthy but simple algebra, from \Ref{ufus}, \Ref{bopt}, \Ref{uopt}
and \Ref{Omega} one can compute: 
\begin{align}
\Tr[U_{\mathrm{opt}}(\tau_\ast)\da U_f H_{\mathrm{opt}}(\tau_\ast)]=2i(-1)^{m+1}\omega_K^2R(\tau_\ast),
 \label{eqR}
\end{align}
where we have defined:
\begin{align}
R(\tau_\ast)&:= [\cos^2\phi +\frac{b_1}{2}(b_1-b_4+2\sin\phi)]s_1
\nonumber
\\
&+[\cos^2\phi +\frac{b_2}{2}(b_2-b_3+2\sin\phi)]s_2
\nonumber
\\
&+[\cos^2\phi -\frac{b_3}{2}(b_2-b_3-2\sin\phi)]s_3
\nonumber
\\
&-[\cos^2\phi -\frac{b_4}{2}(b_1-b_4-2\sin\phi)]s_4.
 \label{R}
\end{align}
Finally, substituting \Ref{m1modsign} and \Ref{eqR} into \Ref{lambda}, we get:
\begin{align}
\lambda=\frac{\mathrm{sign}[M(\tau_\ast)]}{4f \omega_K^2R(\tau_\ast)}.
 \label{lambdaopt}
\end{align}
Thus, if we can determine the values of $M(\tau_\ast)$ and $R(\tau_\ast)$, from the latter equation we have
determined the optimal value of the Lagrange multiplier $\lambda$ for a given fidelity. 
Furthermore, using \Ref{lambdaopt} the multiplier $\lambda$ can be eliminated from \Ref{lmn}-\Ref{lmp} which, after some
simple manipulations, become eqs. \Ref{popt}-\Ref{qopt}, where
\begin{align}
Q(\tau_\ast)&:=\omega_K[(1+K)(s_1+s_4)+(1-K)(s_2-s_3)].
 \label{Q}
\end{align}

\subsection{Perturbative Expansion around $K=1$}

We first compute the function $f_-$ for the case when the target is the entangler gate $U_s^{13}$, i.e. when the
integers are $n_1, n_2=n_1+2q-1, n_3=n_1 +2s-1$ and $n_4=n_1+2r-1$. From its definition, eq. \Ref{fg}, we get:
\begin{align}
f_-&= 2n_1[2(r-q-s)+1]
\nonumber \\
&+4[r(r-1)-q(q-1)-s(s-1)]-1.
 \label{fm}
\end{align}
Now let us assume that the Ising couplings are not equal, i.e. that $|K|\not =1$, and let us compute $(\omega_1\tau_\ast)^2$, $(\omega_2\tau_\ast)^2$, $(\omega_3\tau_\ast)^2$ and $(\omega_4\tau_\ast)^2$.
From the definitions \Ref{bi},\Ref{omegak} and eqs.\Ref{omegaopt} and \Ref{fnot1} we obtain:
\begin{align}
(\omega_K\tau_\ast)^2 &+ 2(\omega_K\tau_\ast)\sin\phi [(1+K)\tau_\ast -m\pi ]
\nonumber \\
&+ [(1+K)\tau_\ast- m\pi ] ^2=(\pi n_1 +\sigma_1)^2,
\label{19a}
\\
(\omega_K\tau_\ast)^2 &+ 2(\omega_K\tau_\ast)\sin\phi [(1-K)\tau_\ast -m\pi ]
\nonumber \\
&+ [(1-K)\tau_\ast- m\pi ] ^2=(\pi n_2 +\sigma_2)^2,
\label{19b}
\\
(\omega_K\tau_\ast)^2 &-2(\omega_K\tau_\ast)\sin\phi [(1-K)\tau_\ast +m\pi ]
\nonumber \\
&+ [(1-K)\tau_\ast +m\pi ] ^2=(\pi n_3 +\sigma_3)^2,
\label{19c}
\\
(\omega_K\tau_\ast)^2 & -2(\omega_K\tau_\ast)\sin\phi [(1+K)\tau_\ast +m\pi ]
\nonumber \\
&+ [(1+K)\tau_\ast +m\pi ] ^2=(\pi n_4 +\sigma_4)^2.
 \label{19d}
\end{align}

Then, summing \Ref{19a} to \Ref{19d} and subtracting from this \Ref{19b} and \Ref{19c}, we obtain 
the duration time of the unitary evolution:
\begin{align}
\tau_\ast \simeq \pi\sqrt{\frac{f_-}{8K}}\left [1+\frac{2 ~f_-^\sigma}{\pi ~f_-}\right ],
 \label{tausigma}
\end{align}
where we have introduced (for notational consistency with \Ref{fg}) $f_\pm^\sigma :=(n_1\sigma_1+n_4\sigma_4)\pm(n_2\sigma_3 +n_3\sigma_3)$, and we discarded contributions of $O(\sigma^2)$.
Subtracting \Ref{19c} from \Ref{19b}, we obtain the constant component $\hat B_z$ of the magnetic field: 
 \begin{align}
\omega_K \sin\phi\simeq \frac{1}{\tau_\ast}\left \{\frac{\pi^2\Delta g }{8(1-K)\tau_\ast}\left [1+\frac{2\Delta g^\sigma}{\pi\Delta g}\right ]+m\pi\right \},
 \label{Bzsigma}
\end{align}
where $g_\pm^\sigma :=(n_1\sigma_1-n_4\sigma_4)\pm (n_2\sigma_3 -n_3\sigma_3)$,
$\Delta g^\sigma:=g_+^\sigma -g_-^\sigma$ (contributions of $O(\sigma^2)$ discarded).
From the sum of \Ref{19a}, \Ref{19b}, \Ref{19c} and \Ref{19d}, we get instead a constraint on $\hat \omega^2$
(and therefore, via \Ref{normHk}, on $\omega_K^2$):
\begin{align}
\hat \omega^2\simeq \frac{\pi^2}{4\tau_\ast^2}\left [f_+ +4m^2 +\frac{m\pi \Delta g }{2\tau_\ast} +\frac{2 f_+^\sigma}{\pi}\right ].
 \label{omegasigma}
\end{align}
Finally, from the sum of \Ref{19a} and \Ref{19c}, to which we subtract \Ref{19b} and \Ref{19d}, we obtain the constraint:
\begin{align}
(1-K)&(g_++g_-)-(1+K)\Delta g
\nonumber \\
&=\frac{2}{\pi}[(1+K)\Delta g-(1-K)(g_+^\sigma+g_-^\sigma)].
 \label{Ksigma}
\end{align}
(modulo $O(\sigma^2)$ contributions).
The last equation explains the origin of the unexpected result of Section IV.
In fact,  for $|K|\not = 1$, the only way for which eq. \Ref{Ksigma} can hold to zero order in $\sigma$ is that  
the constraint $K=g_-/g_+$ must hold. When one is looking for the minimum (odd integer) value of of the function
$f_-$  (i.e., from \Ref{tausigma} the minimum time duration $\tau_\ast$ of the unitary evolution) 
this constraint together with those coming from $(\hat B_0)^2 > 0$ and $\omega_K^2 > 0$ 
(see the discussion below eqs. \Ref{tau}-\Ref{omk}) must be simultaneously satisfied by the integer numbers $n_i$, $q,r$ and $s$. This is a very strong set of constraints.
No surprise that, at least for $f_-$ up to $7$ (we did not try higher, unrealistic values of $f_-$ and, therefore, of 
$\tau_\ast$), only the choice of $n_2=n_3$ is allowed, thus enforcing $g_+=g_-$ and $K=1$.
The only possible way out of this "empasse" is to assume that we can work with $|K|$ close to one and hope that the right hand side and the left hand side of eq. \Ref{Ksigma} are of the same order.    

Let us then assume, for example, that $K=1+\delta K$, where we take $|\delta K|\ll 1$.
First of all, performing an expansion in $\delta K$ of the condition \Ref{Ksigma}, we indeed find that this separates into two
conditions, coming from the zero order and the first order terms in $\delta K$ and $\sigma_i$, i.e., respectively:
 \begin{align}
n_2&=n_3,
\\
n_2(\sigma_2-\sigma_3)&=-\frac{\pi}{4}(n_1^2-n_4^2)\delta K.
 \label{KsigmaK}
\end{align}
That is to say, $\delta K$ must be of the order of $\sigma_i$, i.e., from \Ref{22}, we must have 
$\delta K\lessapprox \epsilon$.

Then, we perform a  perturbative expansion in $\delta K$
of the conditions on the variables $\sigma_i$, i.e. an expansion of eqs. \Ref{13}-\Ref{15}. 
Exploiting \Ref{bi} we obtain, up to the lowest order in both $\delta K$ and $\sigma_i$: 
\begin{align}
\biggl [\omega_K\sin\phi -\frac{\hat \Omega}{2}+2\biggr ]\frac{\sigma_1}{n_1}
=\left [\omega_K\sin\phi -\frac{\hat \Omega}{2}-2\right ]\frac{\sigma_4}{n_4}
 \label{13K}
\\
\left (\frac{\sigma_1}{n_1}-\frac{\sigma_4}{n_4}\right )
=\frac{\hat \Omega}{4}\left (\frac{\sigma_1}{n_1}+\frac{\sigma_2}{n_2}+\frac{\sigma_3}{n_3}+\frac{\sigma_4}{n_4}\right ).
\label{15K} 
\end{align}
We can now solve eqs. \Ref{13K}-\Ref{15K} together with \Ref{22}.
Using the lowest order approximations to the time-optimal evolution 
(eqs.\Ref{tauopt}-\Ref{Bzopt}, or eqs. \Ref{tausigma}-\Ref{Bzsigma}, with $n_1=2, n_2=n_3=n_4=1$ and $f_-=3$), 
we can find that $\omega_K\sin\phi -\hat \Omega/2 \simeq -1$. Therefore, eqs. \Ref{13K}-\Ref{15K} give:
\begin{align}
\sigma_2&=\frac{1}{2}\left [\left (1+\frac{\sqrt{6}}{m}\right )\sigma_1-\frac{3\pi}{4}\delta K\right ],
\\
\sigma_3&=\frac{1}{2}\left [\left (1+\frac{\sqrt{6}}{m}\right )\sigma_1+\frac{3\pi}{4}\delta K\right  ],
\\
\sigma_4&=-\frac{3}{2}\sigma_1.
 \label{sigma234}
\end{align}
Finally, substituting \Ref{sigma234} into \Ref{22}, we obtain:
\begin{align}
\sigma_1=\pm \sqrt{\frac{2\Delta_{\epsilon K}}{\left (1+\frac{\sqrt{6}}{m}\right )^2+\frac{13}{2}}},
 \label{sigma1}
\end{align}
where $\Delta_{\epsilon K}:=\epsilon^2-\frac{9\pi^2}{32}(\delta K)^2$.
Here we notice that, in order for the time-optimal evolution to exist, i.e. for the $\sigma_i$ to be real, we have a more precise 
condition that must be satisfied by the couplings in the Ising Hamiltonian, i.e., we must have that 
$\Delta_{\epsilon K}\geq 0$ or, in terms of the fidelity, using eq. \Ref{fepsilon}:
\begin{align}
|\Delta K|\leq \frac{16}{3\pi}\sqrt{1-f}.
 \label{Kepsilon}
\end{align}
We can now proceed in determining the perturbative expansions in $\delta K$ of the time-optimal 
solutions \Ref{tausigma}-\Ref{Ksigma} for $f_-=3$, which explicitly read (to the lowest order in $\delta K$ and $\sigma_i$):
\begin{align}
\tau_\ast &\simeq \pi\sqrt{\frac{3}{8}}\left (1-\frac{\delta_{K \sigma}}{2}\right ),&
 \label{tausigmaK}
\\
\hat \Omega &\simeq m\sqrt{\frac{8}{3}}\left (1+\frac{\delta_{K \sigma}}{2}\right ),&
\label{OmegasigmaK}
\\
\omega_K \sin\phi &\simeq \left (1+\sqrt{\frac{8}{3}}\right )\biggl [1+\frac{(\sqrt{2}+\sqrt{3})}{(2\sqrt{2}+\sqrt{3})}
\delta_{K \sigma}\biggr ],
\label{BzsigmaK}
\end{align}
where we have defined: 
\begin{align}
\delta_{K \sigma}:=\delta K -\frac{2f_-^\sigma}{3\pi}.
 \label{deltaksigma}
\end{align}
Finally, the value of the integer $m$ is also fixed by minimizing $\tau_\ast$ in \Ref{tausigmaK}, i.e. by minimizing
\begin{align}
f_-^\sigma=-\frac{1}{2}\left(1+\frac{2\sqrt{6}}{m}\right )\sigma_1 
\label{fmsigma}
\end{align}
with $\sigma_1$ given by \Ref{sigma1}, with respect to $m$ (and the sign of $\sigma_1$ in \Ref{sigma1}).
The result is $m=1$ and $\mathrm{sign}(\sigma_1)=+$, for which 
$f_-^\sigma=-(1+2\sqrt{6})[\Delta_{\epsilon K}/(27+4\sqrt{6})]^{1/2}$.
This gives the final formulas \Ref{opttau}-\Ref{optBz} of Section V.

A similar computation can be done for the case of the target $\mathrm{CNOT}(1,3)$, eq. \Ref{uf1}, and for the "rotated" Hamiltonian \Ref{ising2}, discussed in Section VI.
In particular, one has to choose 
$n_2^\prime=n_1^\prime+2q^\prime, n_3^\prime=n_1^\prime+2s^\prime, n_4^\prime=n_1^\prime+2r^\prime +1$,
with $n_1^\prime, n_2^\prime, n_3^\prime$ and $n_4^\prime$ positive integers, and $q^\prime, s^\prime$ and $r^\prime$ integers.
One obtains, for $n_1=2$ and $n_2=n_3=n_4=1$, the same time-optimal solutions \Ref{opttau}-\Ref{optomegak}, except for a change everywhere in in the sign in front of odd powers of the integer $m$, due to the fact that now $m=-1$ is the value minimizing $\tau_\ast$.

\newpage

\vspace{1cm}

\section*{ACKNOWLEDGEMENTS}
This research was partially supported by the MEXT of Japan, 
 under grant No.  (T.K.) and by the project Rientro dei Cervelli of
 the MIUR of Italy (A.C.).

\bibliographystyle{alpha}

\end{document}